\input harvmac
\input epsf
\noblackbox
\newcount\figno
\figno=0 
\def\fig#1#2#3{
\par\begingroup\parindent=0pt\leftskip=1cm\rightskip=1cm\parindent=0pt
\baselineskip=11pt
\global\advance\figno by 1
\midinsert
\epsfxsize=#3
\centerline{\epsfbox{#2}}
\vskip 12pt
{\bf Fig.\ \the\figno: } #1\par
\endinsert\endgroup\par
}
\def\figlabel#1{\xdef#1{\the\figno}}

\skip0=\baselineskip
\divide\skip0 by 2
\def\p{\partial}
\def\ds{dS$_3$}
\def\hi{$\hat { \cal I}^-$}
\def\im{$ { \cal I}^-$}
\def\ip{$ { \cal I}^+$}
\def\ri{{\rm in}}
\def\ro{{\rm out}}
\def\msurr{\mathsurround=0pt}
\def\overleftrightarrow#1{\vbox{\msurr\ialign{##\crcr
        $\leftrightarrow$\crcr\noalign{\kern-1pt\nointerlineskip}
        $\hfil\displaystyle{#1}\hfil$\crcr}}}

\lref\coss{
O.~Coussaert, M.~Henneaux and P.~van Driel,
``The Asymptotic dynamics of three-dimensional Einstein gravity with a negative cosmological constant,''
Class.\ Quant.\ Grav.\  {\bf 12}, 2961 (1995)
[gr-qc/9506019].}
\lref\pol{J. Polchinski, {\it String Theory} Vol. 1, Cambridge University 
Press (1998).}
\lref\raf{R. Bousso, private comunication.}
\lref\hms{
S.~Hawking, J.~Maldacena and A.~Strominger,
``DeSitter entropy, quantum entanglement and AdS/CFT,''
JHEP {\bf 0105}, 001 (2001)
[hep-th/0002145].}
\lref\bousso{
R.~Bousso,
``Bekenstein bounds in de Sitter and flat space,''
JHEP {\bf 0104}, 035 (2001)
[hep-th/0012052].}
\lref\hulla{
C.~M.~Hull,
``Timelike T-duality, de Sitter space, large N gauge theories and  topological field theory,''
JHEP {\bf 9807}, 021 (1998)
[hep-th/9806146].}
\lref\hullb{
C.~M.~Hull and R.~R.~Khuri,
``Worldvolume theories, holography, duality and time,''
Nucl.\ Phys.\ B {\bf 575}, 231 (2000)
[hep-th/9911082].}
\lref\juan{J.~Maldacena,
``The large N limit of superconformal field theories and supergravity,''
Adv.\ Theor.\ Math.\ Phys.\  {\bf 2}, 231 (1998)
[hep-th/9711200].}
\lref\jmas{
J.~Maldacena and A.~Strominger,
``Statistical entropy of de Sitter space,''
JHEP {\bf 9802}, 014 (1998)
[gr-qc/9801096].}
\lref\asbh{A.~Strominger,
``Black hole entropy from near-horizon microstates,''
JHEP {\bf 9802}, 009 (1998)
[hep-th/9712251].}
\lref\rbas{R. Bousso and A. Strominger, in preparation.}
\lref\by{
J.~D.~Brown and J.~W.~York,
``Quasilocal energy and conserved charges derived from the gravitational action,''
Phys.\ Rev.\ D {\bf 47}, 1407 (1993).}
\lref\bk{V.~Balasubramanian and P.~Kraus,
``A stress tensor for anti-de Sitter gravity,''
Commun.\ Math.\ Phys.\  {\bf 208}, 413 (1999)
[hep-th/9902121].}
\lref\bh{
J.~D.~Brown and M.~Henneaux,
``Central Charges In The Canonical Realization Of Asymptotic Symmetries: An Example From Three-Dimensional Gravity,''
Commun.\ Math.\ Phys.\  {\bf 104}, 207 (1986).}
 \lref\witt{E.~Witten,
``Anti-de Sitter space and holography,''
Adv.\ Theor.\ Math.\ Phys.\  {\bf 2}, 253 (1998)
[hep-th/9802150].}
\lref\hs{M.~Henningson and K.~Skenderis,
``The holographic Weyl anomaly,''
JHEP {\bf 9807}, 023 (1998)
[hep-th/9806087].} 
\lref\ad{L.~F.~Abbott and S.~Deser,
``Stability Of Gravity With A Cosmological Constant,''
Nucl.\ Phys.\ B {\bf 195}, 76 (1982).} 
\lref\desit{ Some particularly useful references are E.~Mottola,
``Particle Creation In De Sitter Space,''
Phys.\ Rev.\ D {\bf 31}, 754 (1985); B.~Allen,
``Vacuum States In De Sitter Space,''
Phys.\ Rev.\ D {\bf 32}, 3136 (1985); N. D. Birrel and P. C. W. 
Davies, {\it Quantum Fields in Curved Space}, Cambridge University Press, 
Cambridge (1982).   }
\lref\sei{N.~Seiberg,
``Notes On Quantum Liouville Theory And Quantum Gravity,''
Prog.\ Theor.\ Phys.\ Suppl.\  {\bf 102}, 319 (1990).}
\lref\giddings{S.~B.~Giddings,
``The boundary S-matrix and the AdS to CFT dictionary,''
Phys.\ Rev.\ Lett.\  {\bf 83}, 2707 (1999)
[hep-th/9903048].}
\lref\ascv{
A.~Strominger and C.~Vafa,
``Microscopic Origin of the Bekenstein-Hawking Entropy,''
Phys.\ Lett.\ B {\bf 379}, 99 (1996)
[hep-th/9601029].}
\lref\bek{
J.~D.~Bekenstein,
``Black Holes And Entropy,''
Phys.\ Rev.\ D {\bf 7}, 2333 (1973).}
\lref\banksb{
T.~Banks,
``Cosmological breaking of supersymmetry or little Lambda goes back to  the future. II,''
hep-th/0007146.}
\lref\boussoc{
R.~Bousso,
``Holography in general space-times,''
JHEP {\bf 9906}, 028 (1999)
[hep-th/9906022].}
\lref\hawk{S.~W.~Hawking,
``Particle Creation By Black Holes,''
Commun.\ Math.\ Phys.\  {\bf 43}, 199 (1975).}
\lref\vijay{
V.~Balasubramanian, P.~Horava and D.~Minic,
``Deconstructing de Sitter,''
JHEP {\bf 0105}, 043 (2001)
[hep-th/0103171].}
\lref\gibbons{
G.~W.~Gibbons and S.~W.~Hawking,
``Cosmological Event Horizons, Thermodynamics, And Particle Creation,''
Phys.\ Rev.\ D {\bf 15}, 2738 (1977).}
\lref\nastya{
A.~Volovich,
``Discreteness in deSitter space and quantization of Kahler manifolds,''
hep-th/0101176.}
\lref\britto{
R.~Britto-Pacumio, A.~Strominger and A.~Volovich,
``Holography for coset spaces,''
JHEP {\bf 9911}, 013 (1999)
[hep-th/9905211].}
\lref\at{
A.~Achucarro and P.~K.~Townsend,
``A Chern-Simons Action For Three-
Dimensional Anti-De Sitter Supergravity Theories,''
Phys.\ Lett.\ B {\bf 180}, 89 (1986).}
\lref\wu{
F.~Lin and Y.~Wu,
``Near-horizon Virasoro symmetry and the entropy of de 
Sitter space in  any dimension,''
Phys.\ Lett.\ B {\bf 453}, 222 (1999)
[hep-th/9901147].}
\lref\kim{
W.~T.~Kim,
``Entropy of 2+1 dimensional de Sitter space in terms of brick wall  method,''
Phys.\ Rev.\ D {\bf 59}, 047503 (1999)
[hep-th/9810169].}
\lref\banados{
M.~Banados, T.~Brotz and M.~E.~Ortiz,
``Quantum three-dimensional de Sitter space,''
Phys.\ Rev.\ D {\bf 59}, 046002 (1999)
[hep-th/9807216].}
\lref\gao{
S.~Gao and R.~M.~Wald,
``Theorems on gravitational time delay and related issues,''
Class.\ Quant.\ Grav.\  {\bf 17}, 4999 (2000)
[gr-qc/0007021].}
\lref\cc{See e.g. S.~Perlmutter,
``Supernovae, dark energy, and the accelerating universe: The status of  the cosmological parameters,''
in {\it Proc. of the 19th Intl. Symp. on Photon and Lepton Interactions at High Energy LP99 } ed. J.A. Jaros and M.E. Peskin,
Int.\ J.\ Mod.\ Phys.\ A {\bf 15S1}, 715 (2000).}
\lref\boussob{
R.~Bousso,
``Positive vacuum energy and the N-bound,''
JHEP {\bf 0011}, 038 (2000)
[hep-th/0010252].}
\lref\witcs{
E.~Witten,
``(2+1)-Dimensional Gravity As An Exactly Soluble System,''
Nucl.\ Phys.\ B {\bf 311}, 46 (1988).}
\lref\witb{
E.~Witten,
``Quantization of Chern Simon Theory with Complex Gauge Group,''
Commun.\ Math.\ Phys.\  {\bf 137}, 29 (1991).}
\lref\banks{
T.~Banks and W.~Fischler,
``M-theory observables for cosmological space-times,''
hep-th/0102077.}
\lref\witc{
E.~Witten, ``Quantum gravity in de Sitter space'' Strings 2001 online 
proceedings http://theory.theory.tifr.res.in/strings/Proceedings }
\Title{\vbox{\baselineskip12pt\hbox{hep-th/0106113}\hbox{}
\hbox{}}}{The dS/CFT Correspondence}

\centerline{Andrew Strominger }
\bigskip\centerline{Department of Physics}
\centerline{Harvard University}
\centerline{Cambridge, MA 02138}

\vskip .3in \centerline{\bf Abstract}
A holographic duality is proposed relating quantum gravity on 
dS$_D$ (D-dimensional de Sitter space) to conformal field theory on 
a single 
S$^{D-1}$ ((D-1)-sphere), in which bulk de Sitter correlators with 
points on the boundary are related to CFT correlators on the sphere,
and 
points on \ip\ (the future  boundary of dS$_D$) are mapped to 
the antipodal points on S$^{D-1}$ relative to those on \im.  
For the case 
of \ds , which is analyzed in some detail, the central charge of the 
CFT$_2$ is 
computed in an analysis of  the asymptotic
symmetry group at ${\cal I}^\pm$. 
This dS/CFT proposal is supported by the computation of correlation 
functions of a massive 
scalar field. 
In general the 
dual CFT may be non-unitary and (if for example 
there are sufficently massive stable scalars) contain complex conformal
weights. We also consider 
the physical region ${\cal O}^-$ of \ds\ corresponding to the causal past of a timelike
observer, whose holographic dual lives on a plane rather than a sphere. 
${\cal O}^-$ can be foliated by 
asymptotically flat spacelike slices. Time evolution along these slices is 
generated by $L_0+\bar L_0$, and is dual to scale transformations 
in the boundary CFT$_2$.
 
\smallskip
\Date{}
\listtoc
\writetoc
\newsec{Introduction}

  The macroscopic entropy-area law \refs{\bek, \hawk}
\eqn\aren{S={A \over 4 G}.}
relates thermodynamic entropy to the area of an event horizon. A striking
feature of this law is its 
universal applicability, including all varieties of black holes
as well as de Sitter \gibbons\  and Rindler spacetimes. Understanding the 
microscopic origin of \aren\ is undoubtedly a key step towards 
understanding the fundamental nature of spacetime and quantum mechanics. 
Some progress has recently been made in deriving \aren\ for certain black
holes in string theory \ascv. This has led to a variety of insights 
culminating in the AdS/CFT correspondence \juan. However the 
situation remains unsatisfactory in that these recent developments do 
not fully explain the universality of \aren. 

In particular one would like to derive the entropy and 
thermodynamic properties of 
de Sitter space. This has taken on added significance with the emerging
possibility that the real universe resembles de Sitter space \cc. Recent 
discussions of de Sitter thermodynamics   
include\refs{\jmas  \hulla \banados \kim \wu \boussoc  \hullb \hms \banksb  
\bousso \witc 
\nastya \banks -\vijay  }.  
An obvious approach, successfuly employed in the black
hole 
case,  would be to begin by embedding de Sitter space as 
a solution of string theory, and then exploit various string dualities to 
obtain a microscopic description. Unfortunately persistent 
efforts by many (mostly unpublished!) have so far failed even to find a 
fully satisfactory de Sitter solution of string theory.  Hopefully 
this situation will change in the not-too-distant future. 

Meanwhile, string theory  may not be the only route to at least a 
partial understanding 
of de Sitter space. Recall that the 
dual relation between AdS$_3$ and CFT$_2$ was discovered by 
Brown and Henneaux \bh\ from a general analysis 
of the asymptotic symmetries of anti-de Sitter space, and the central charge 
of the CFT$_2$ was computed. Later on black hole entropy was 
derived \asbh\ using this central charge and Cardy's formula. In principle
this required no input from string theory. Of course the arguments of 
\asbh\ would have been less convincing without the concrete examples
supplied by string theory. 

In the absence of a stringy example of de Sitter space, in this paper 
we will sketch the parallel steps, beginning with an analysis of the
asymptotic symmetries along the lines of \bh,  
toward an understanding of de Sitter space. The endpoint will be a holographic
duality relating quantum gravity on de Sitter space to a euclidean CFT on a 
sphere of one lower dimension.  
Our steps will be guided by the analogy to the 
the AdS/CFT correspondence. We will see many similarities but also 
important differences between the AdS/CFT and 
proposed dS/CFT correspondences.

One of the first issues that must be faced in discussions of 
quantum de Sitter is the spacetime region under consideration. 
An initial reaction might be to consider the entire 
spacetime, which contains two boundaries ${\cal I}^\pm$ which are past and
future 
spheres. This may ultimately be the correct view, but it 
is problematic for several reasons. 
The first is that a single immortal 
observer in de Sitter space can see at most half of the space. 
So a description of the entire space goes beyond what can 
be physically measured. Trying to describe the entirety of de Sitter 
space is like trying to describe the inside and outside of a 
black hole at the same time, and may lead to trouble. A second problem 
recently stressed in \boussob\
is that, if enough matter is present, a space which is 
asymptotically de Sitter at \im\ may collapse at finite time, and there
will be no future \ip\ de Sitter region at all. Even when collapse 
does not occur, 
the presence of matter alters the causal structure \gao. This 
obscures the relevance of the global de Sitter geometry. 

A smaller region\foot{References \refs{\jmas,\wu ,\hms } advocate an even
smaller causal 
region corresponding to the interior of both the past and future horizons
of a timelike observer. This even smaller region excludes both \im\ 
and \ip.}, denoted ${\cal O}^-$ herein, is the region which
can be seen be a single timelike observer in de Sitter space. It includes 
the planar 
past asymptotic region \hi , which is \im\ minus a point, but not \ip. 
Discussion of the quantum physics of ${\cal O}^-$ does not include
unobservable regions and does not presume the existence of \ip. 
In this
paper we will consider the holographic duals for both this region and the 
full space. One of our conclusions will be that in both cases the dual 
is a single euclidean CFT, despite the two boundaries of the full space,  
so from the dual perspective the two cases 
are not as different as they might seem.

We begin in section 2 with a discussion of the asymptotic symmetries of 
\ds.  It is shown that conformal diffeomorphisms of the spacelike surfaces
can be compensated for by shifts in the time coordinate in such a way that 
the asymptotic form of the 
metric on \hi\ is unchanged.  This is similar to the 
AdS$_3$ case \bh\ except that time and radial coordinates are exchanged. 
Hence the asymptotic symmetry group of \ds\ is the euclidean conformal 
group in two dimensions. The global $SL(2,C)$ subgroup is the \ds\ isometry
group. In section 3 we introduce the Brown-York stress tensor \by\ for the 
boundary of \ds. In section 4 we determine the central charge of the CFT
(following \bk) from the anomalous variation of this boundary stress
tensor.
We find $c={3 \ell \over 2 G}$, with $\ell$ the de Sitter radius and $G$ 
Newton's constant.  In section 5 we study correlators for massive scalar
fields with points on \hi. It is found that they have the right form to be
dual to CFT correlators of conformal fields on the plane, 
with conformal weights
determined from the scalar mass. For $m^2\ell^2>1$, the conformal 
weights become complex. This means that the boundary CFT is not unitary 
if there are stable scalars with masses above this bound. 

In section 
6 we turn to global \ds\  which has  two asymptotic S$^2$ regions.
We first show that scalar correlators with points only on \im\ are dual to CFT 
correlators on the sphere.  We then consider the case with one point on 
\im\ and one point on \ip.  These have singularities when the point on \ip\
is antipodal to the point on \im.  This is because a light ray beginning 
on the sphere at \im\ reaches its antipode at \ip, and so antipodal points are connected
by null geodesics. This  causal connection relating points on \im\ to those
on \ip\ breaks the two copies of the conformal group (one each for 
${\cal I}^\pm$) down to a single copy. After inverting the argument of 
the boundary field on \ip, correlators with one point on \ip\ and one point
on \im\ have the same form as those with both points on \im\ (or both on
\ip). Hence we propose that the dual CFT lives on a single euclidean 
sphere, rather than 
two spheres as naively suggested by the nature of the boundary of global \ds.
\foot{ Since according to this proposal, 
the boundary CFT lives on a sphere in any case, the
question of whether or not there is an \ip\ at all might then be rephrased
in terms of properties of the CFT.} 

In section 7 we return to the region ${\cal O}^-$ of \ds. This region can be
foliated 
by asymptotically flat spacelike slices. Quantum states can be defined on
these slices and at \hi\ form a representation of the conformal group. 
Time evolution is generated by $L_0+\bar L_0$. In the dual CFT this is
scale transformations. This is the de Sitter analog of the 
scale-radius duality in AdS/CFT, and may have interesting implications for
cosmology. We close in section 8 with a brief discussion of generalizations 
to higher than three dimensions.  An appendix includes some details of \ds\ 
geometry and Green functions.

It will be evident to the reader that our understanding of the proposed
dS/CFT correspondence is incomplete.  A complete understanding will
ultimately require an example of de Sitter quantum gravity. 
We hope that the present work will help guide us
in what to look for. 

The closest things we have at present to examples are Hull's
spacelike D-branes \refs{\hulla, \hullb} and  
Chern-Simon de Sitter gravity \refs{\at,\witcs}.  Hull performs 
timelike T-duality to turn for example AdS$_5\times $S$_5$ to 
into dS$_5\times$ H$_5$, where H$_5$ is the hyperbolic 5-plane, and argues
that the dual is a spacelike 3-brane. As discussed in  \refs{\hulla,
\hullb},  this example is pathological because there are fields with the
wrong sign kinetic term. Nevertheless it may be an instructive example for
some purposes. 
Alternately, 
pure gravity in de Sitter space can be written as an $SL(2,C)$ Chern-Simon 
gauge theory, which is holographically dual to a reduced\foot{As in 
the AdS case \coss , we expect this becomes Liouville theory after imposing the
appropriate boundary condations. Some discussion can be found in 
\banados.}  $SL(2,C)$ WZW model on 
the boundary. This latter theory (before reduction) 
was studied extensively in \witb\ as 
the complexification of $SU(2)$, but its status remains unclear. 
Both of these examples deserve further exploration.  

The notion that quantum gravity in de Sitter space may have a 
euclidean holographic dual, possibly related to \im\ and/or \ip , has arisen 
in a number of places, including 
\refs{\hulla, \boussoc,\witc, \vijay}.
\newsec{Asymptotic Symmetries of \ds\ }
The region ${\cal O}^-$ comprising the causal
past 
of a timelike observer in de Sitter space is illustrated in figure 1. 
\fig{Penrose diagram for \ds. Every point in the interior of the diagram is 
an $S^1$. A horizontal line is an $S^2$, with the left (right) vertical
boundary being the north (south) pole. ${\cal O}^-$ is the region 
below the
diagonal, and comprises the causal past of an observer at the south pole. 
The dashed lines are non-compact surfaces of constant $t$.
}{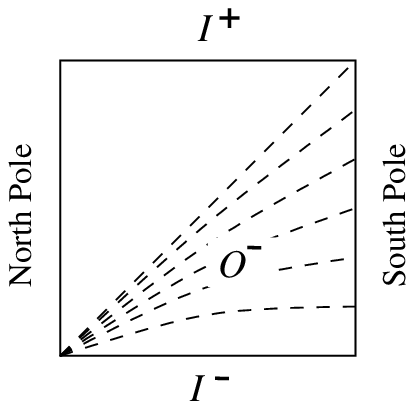}{2in}
The metric for a planar slicing of ${\cal O}^-$ is 
given   by \eqn\dsmet{{ds^2 \over \ell^2}=e^{-2t}dzd\bar z -dt^2.} 
We use \hi\ to denote the plane which is the past infinity of ${\cal O}^-$.  
An asymptotically past de Sitter geometry is one for which the metric 
behaves for $t \to -\infty$ as  
\eqn\tyip{\eqalign{g_{z \bar z}&={e^{-2t} \over 2} +{\cal O}(1),\cr
g_{tt}&=-1 +{\cal O}( e^{2t} ),\cr
g_{zz}&={\cal O}( 1 ),\cr g_{zt}&={\cal O}( e^{3t} ).\cr}}
These boundary conditions are an analytic continuation of 
the $AdS_3$ boundary conditions of Brown and Henneaux \bh.
The asymptotic symmetries of \ds\ are diffeomorphisms 
which preserve \tyip.   
Consider the vector fields
\eqn\dffe{\zeta=U\p_z+{1 \over 2}e^{2t} U^{\prime
\prime} \p_{\bar z}+{1 \over 2} U^{\prime}\p_t,} where $U=U(z)$ is
holomorphic and the prime denotes differentiation. In order to
obtain a  real vector field one must add the complex conjugate,
however for notational simplicity we suppress this addition in
\dffe\ and all subsequent formula. In general the metric 
transforms under a diffeomorphism as the Lie derivative 
\eqn\rll{\delta_\zeta  g_{mn}=-{\cal L}_\zeta g_{mn}.}
For $\zeta$ parametrized by $U$ as in \dffe, \rll\ becomes 
\eqn\ggvc{\eqalign{\delta_U g_{zz}&=-{\ell^2 \over 2} U^{\prime
\prime\prime},\cr \delta_U g_{z\bar z}&=\delta_U g_{zt}=\delta_U g_{tt}=0.}}
The change \ggvc\ in the metric satisfies \tyip\ and so 
\dffe\ generates an asymptotic symmetry of de Sitter
space on \hi . 

A special case of \dffe\ is
\eqn\drfv{U=\alpha +\beta  z +\gamma z^2,}
where $\alpha ,~\beta  ,~\gamma $ are complex constants.
In this case $U^{\prime\prime\prime}$ vanishes, and the metric
is therefore invariant. These transformations generate the
$SL(2,C)$ global isometries of 2+1 de Sitter.

In conclusion the asymptotic symmetry group is the
conformal group of the complex plane, and the isometry group is
$SL(2,C)$ subgroup  of the asymptotic symmetry group.

\newsec{The Boundary Stress Tensor}

Brown and York \by\ have given a general prescription for defining
a stress tensor associated to a boundary of a spacetime. Our
treatment parallels the discussion of AdS$_3$ in  \bk. We will be
interested in the boundary $t\to -\infty$ at \hi . In the
case at hand the stress tensor is\foot{Our conventions in this section are
 those of \refs{\by, \bk}  except for a factor of $-2\pi$ in this equation, 
in order to conform with 
the more standard conventions \pol\ for 2D CFT stress tensor.} \eqn\gty{T^{\mu
\nu}
=-{4 \pi \over
\sqrt{\gamma}}{\delta S \over \delta \gamma_{\mu\nu}},} where
$\gamma $ is the induced metric on the boundary. The action is
\eqn\fed{S={1 \over 16\pi G}\int_{\cal M} d^3x\sqrt{-g}(R-{2 \over
\ell^2}) +{1 \over 8\pi G}\int_{\p \cal M}\sqrt{\gamma}K +{1 \over
8\pi G
 \ell}\int_{\p \cal M}\sqrt{\gamma}+S_{\rm matter},}
$K$ here is the trace of the extrinsic curvature
defined by $K_{\mu\nu}=-\nabla_{(\mu}n_{\nu )}=-\half 
{\cal  L}_n \gamma_{\mu\nu}$ with $n^\mu$ the outward
 pointing unit normal. The second integral in \fed\ is the usual
gravitational surface term. The third integral is a surface
counterterm required for finiteness of $T$ at an asymptotic
boundary, and uniquely fixed by
 locality and  general covariance \bk. The matter action is assumed not
to be relevant near the boundary and will henceforth be suppressed.

The sign in the definition \gty\ 
leads to a positive mass\foot{ More precisely, a positive value of the 
AD mass $L_0+\bar L_0$, as defined below in equation 7.4.} for Schwarzschild-de
Sitter \rbas.  It would be interesting to see if \gty\ reproduces the 
canonical operator product expansion for a 2D CFT stress tensor.

Using \fed\ to evaluate \gty\ we learn that, 
for solutions of the bulk equations
of motion,
\eqn\gtyz{T^{\mu \nu}={1 \over 4 G}\bigl[ K^{\mu\nu}-
(K+{1 \over \ell})\gamma^{\mu\nu} \bigr].} This vanishes for de Sitter
space 
on \hi\ in the
coordinates \dsmet. For more general asymptotically \ds\ spacetimes 
(obeying \tyip) \gtyz\ implies 
\eqn\gtz{T_{zz}={1 \over 4 G}\bigl[ K_{zz}+
{1 \over \ell}\gamma_{zz} \bigr].}

\newsec{The Central Charge}
A central charge can be associated to \ds\  by analyzing the
behavior of the stress tensor on $\cal I^-$. We follow related
discussions for AdS$_3$ \refs{\bh,\bk} 
(which in turn followed the earlier work \hs). Under the conformal
transformations \dffe, one finds \eqn\ctm{\delta_U T_{zz}=-{\ell
\over 8 G}U^{\prime\prime\prime}.} This transformation
identifies the central charge as\foot{This equation was arrived at from a
different perspective in \banados , and a related equation appears in \wu.} 
\eqn\cgh{c={3 \ell \over 2 G}.}

It is presumably also possible to derive $c$ by
an alternate method \refs{\witt, \hs,\bk}
which relates it to the trace anomaly by using spherical rather than
planar spatial sections.

\newsec{The Plane and \hi \ Correlators  }

In the previous sections we have seen that the conformal group of euclidean
$R^2$ has an action on \hi.  One therefore expects that appropriately
rescaled gravity
correlators restricted to \hi\ will be those of a euclidean 2D conformal
field theory. In this section we verify this expectation for the case of a
massive scalar.

  Consider a scalar field of mass $m$ with 
wave
equation \eqn\phie{m^2\ell^2\phi=\ell^2 \nabla^2\phi=
-\p_t^2\phi+{2 }\p_t \phi +4e^{2t} \p_z\p_{\bar z}
\phi.} Near \hi\ the last
term in \phie\ is negligible and solutions behave as \eqn\tgv{\phi
\sim e^{h_\pm t},~~~~~t \to -\infty,} where \eqn\hmrel{h_\pm=1\pm
\sqrt{1-m^2\ell^2}.} We first consider the case $0<m^2\ell^2<1$ so
that $h_\pm$ are real and $h_-<1<h_+$. As a boundary condition on
\hi\ we demand \eqn\efc{\lim_{t \to - \infty} ~~\phi(z, \bar z,
t)=e^{h_- t}\phi_- (z, \bar z),}
with corrections suppressed by at least one power of $e^{2t}$.
The \ds\ /CFT$_2$ correspondence
proposes, in direct analogy with the AdS/CFT correspondence, 
that $\phi_-$ is dual to an operator ${\cal O}_\phi$ of
dimension $h_+$ in the boundary CFT. The two point
correlator of ${\cal O}_\phi$ is (up to normalization)
the quadratic coefficient of $\phi_-$ in the expression\foot{This is
equivalent to the usual expression used in AdS/CFT (except of course
with a different boundary
and $G$), as can be readily seen
from the formula for the bulk Green function $G$ in terms of
the bulk-to-boundary Green function. A clear discussion
can be found in \giddings . We have avoided use of the
 bulk-to-boundary Green function
because in the de Sitter case we need to keep track of both terms in
(5.6).}
\eqn\rtcn{\lim_{t\to -\infty} \int_{\hat {\cal I}^-} d^2zd^2v
\bigl[e^{-2(t+t')} \phi (t,z,
\bar z)\overleftrightarrow \p_t  G(t,z,
\bar z ;t', v, \bar v )\overleftrightarrow \p_{t'}  \phi (t', v, \bar v)
\bigr]_{t=t'}.}
$G$ here is the de Sitter invariant Green function given in the appendix.
Near \hi\ it reduces to
\eqn\geq{ \lim_{t,t' \to - \infty} G(t,z,
\bar z ;t', v, \bar v )={c_+ e^{h_+(t+t')} \over |z-v|^{2h_+}}+
{c_- e^{h_-(t+t')} \over |z-v|^{2h_-}},}
where $c_\pm$ are constants.  Inserting \efc\ and
\geq, \rtcn\ is proportional to\foot{Similar holographic 
expressions were derived for general coset spaces in \britto , 
but the case of de Sitter space was not explicitly considered.}
\eqn\xxj{\int_{\hat {\cal I}^-} d^2zd^2v\phi_- (z,
\bar z)|z-v|^{-2h_+} \phi_- (v, \bar v).}
We conclude that the dual operator ${\cal O}_\phi$ obeys
\eqn\fdnn{\langle {\cal O}_\phi(z,\bar z){\cal O}_\phi (v, \bar v) \rangle
={{\rm const.} \over |z-v|^{2h_+}},}
as is appropriate for an operator of dimension $h_+$. 

It should be noted that the boundary conditions \efc\ are not the most
general. There are also solutions with the subleading behavior
$ \phi_+ (z,\bar z)e^{h_+t}$ at \hi. Including these would lead to an
additional term in \xxj\ proportional to 
\eqn\xj{\int_{\hat {\cal I}^-} d^2zd^2v\phi_+ (z,
\bar z)|z-v|^{-2h_-} \phi_+ (v, \bar v),} which might be
associated with an operator of dimension $h_-$. In
the next section we will find that this extra boundary condition can 
imposed on the second boundary  at \ip,
which is not within the coordinate patch ${\cal O}^-$ covered in
the planar coordinates \dsmet. Alternately this second set of independent 
fields might
be eliminated by imposing a suitable boundary condition at the future
horizon 
$t\to \infty$, which would lead to a different Green function.\foot{I thank 
J. Maldacena for this suggestion.}

What happens if $m^2\ell^2>1$? In that case the conformal weight
$h_-$ is no longer real. $h_\pm$ are complex conjugates with real
part equal to unity. Of course the appearance of imaginary
conformal weights suggests that the dual CFT is not unitary. This
might mean that consistent theories of de Sitter quantum gravity
have no stable scalars with $m^2\ell^2>1$. On the other hand we know of
no obvious reason that the dual CFT needs to be unitary.

The preceding discussion is reminiscent of Liouville theory and
indeed suggests that the boundary CFT has a Liouville-like form.
In some discussions of Liouville theory operators with complex
dimensions are encountered. Further there are various kinds of
operators (called macroscopic and microscopic in \sei) which may
or may not be allowed depending on the context.

\newsec{The Sphere and ${\cal I}^\pm$ Correlators}

An alternate form of the \ds\ metric is \eqn\spr{{ds^2\over\ell^2}
=-d\tau^2+4 \cosh^2{\tau } {dwd\bar w \over (1+w\bar w)^2}}
$w=\tan {\theta \over 2} e^{i\phi}$ here is a complex coordinate
on the round sphere. This metric describes \ds\ as a
contracting/expanding two-sphere. These coordinates cover the
entire space which has future and past $S^2$ boundaries ${\cal
I}^\pm$, as depicted in figure 2. 
In general we can consider correlators with points on
either or both of the boundaries.

\fig{Lines of constant $\tau$ in global spherical coordinates are 
the spacelike two-spheres indicated by dashed lines.  
}{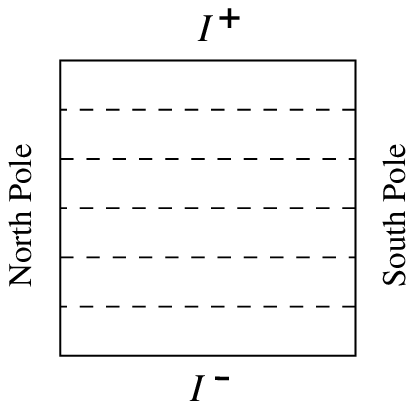}{2in}

We begin with the two-point correlator with both points on \im. We
wish to compute the spherical analog of \rtcn, which is
\eqn\rcn{\lim_{\tau\to -\infty} \int_{ {\cal I}^-} {d^2wd^2v
\sqrt{h(w)h(v)}} \bigl[e^{-2(\tau+\tau')} \phi (\tau,w, \bar
w)\overleftrightarrow \p_\tau  G(\tau,w, \bar w ;\tau', v, \bar
v )\overleftrightarrow \p_{\tau'} \phi (\tau', v, \bar v)
\bigr]_{\tau=\tau'},} where $h(w)=2(1+w\bar w)^{-2}$ is the measure
on the sphere. Near \im, $\phi$ can be decomposed as
\eqn\spq{\lim_{\tau \to -\infty}\phi(\tau, w, \bar w)= \phi_+^\ri (w, \bar w)
e^{h_+\tau}+\phi_-^\ri (w, \bar w) e^{h_-\tau}.} The superscripts "in" ("out")
are used to denote quantities on \im\ (\ip). As seen in the
appendix, the propagator behaves as \eqn\ftl{\eqalign{\lim_{\tau,
\tau' \to -\infty}G(\tau,w, \bar w ;\tau', v, \bar
v)&=c_+e^{h_+(\tau+\tau')} {(1+w\bar w)^{h_+}(1+v\bar v)^{h_+} \over
|w-v|^{2h_+}}\cr &~~~+c_-e^{h_-(\tau+\tau')} {(1+w\bar
w)^{h_-}(1+v\bar v)^{h_-} \over |w-v|^{2h_-}}.}} \rcn\ is then 
proportional to 
\eqn\rfmk{\eqalign{\int_{{\cal I}^-}
d^2vd^2w \sqrt{h(v)h(w)}\bigl( c_+\phi^\ri_-(w,\bar w)
&\Delta_{h_+}(w, \bar w ;v, \bar v)\phi^\ri_-(v, \bar v)
\cr &+c_-\phi^\ri_+(w, \bar w) \Delta_{h_-}(w, \bar w ;v, \bar
v)\phi^\ri_+(v,\bar v) \bigr).}} $\Delta_{h_\pm}$ here is the two
point function for a conformal field of dimension $h_\pm$ on the sphere:
\eqn\wwd{\Delta_{h_\pm}= \bigr[{ (1+w \bar w)(1+v
\bar v) \over  |w-v|^2} \bigr]^{h_\pm},} 
including the normalization factor from
the Weyl anomaly on a curved geometry. Hence we see that, as in the planar
case, the two-point scalar correlators can be identified with
correlators of conformal fields of dimension $h_\pm$, except that
now they are on the sphere rather than the plane.  A similar
expression holds for the boundary at \ip.

Life becomes more interesting when we put one point on \im\ and
one on \ip. Then we must compute \eqn\rcnd{\lim_{\tau\to -\infty}
\int_{ {\cal I}^-} {d^2w \int _{{\cal I}^+}d^2v \sqrt{h(w)h(v)}}
\bigl[e^{2(\tau-\tau')} \phi (\tau,w, \bar w)\overleftrightarrow
\p_\tau  G(\tau,w, \bar w ;\tau', v, \bar v )\overleftrightarrow
\p_{\tau'} \phi (\tau', v, \bar v) \bigr]_{\tau=-\tau'}.} For this
case, as shown in the appendix, the relevant limit of the Green function is 
\eqn\rtv{\lim_{\tau \to -\infty, \tau' \to
+\infty}G(\tau,w, \bar w ;\tau', v, \bar v)= c_+
\cos (\pi h_+) e^{h_+(\tau-\tau')} \Delta_{h_+}(w,\bar w; -{1 \over \bar v},-{1 \over
v} )+(h_+ \leftrightarrow h_-).} 
It is convenient to define the inverted boundary field 
at \ip
\eqn\gtl{\tilde
\phi_+^\ro (v, \bar v)= \phi_+^\ro (-{1 \over \bar v},-{1 \over v}).}
Then \rcnd\ is proportional to
\eqn\rjfmk{\eqalign{\int_{S^2}
d^2wd^2v \sqrt{h(w)h(v)}\bigl( c_+&\cos (\pi h_+) \phi^\ri_-(w,\bar w)
\Delta_{h_+}(w, \bar w ;v, \bar v)\tilde \phi^\ro_-(v, \bar v)
\cr &+c_-\cos (\pi h_-)\phi^\ri_+ (w, \bar w) \Delta_{h_-}(w, \bar w ;v, \bar
v)\tilde \phi^\ro_+(v,\bar v) \bigr).}}
In particular we see that $\phi^\ri$ and $\tilde \phi^\ro$ have a 
non-trivial two point function despite the fact that they live on widely
separated boundary components.

\gtl\ and \rjfmk\ can be interpreted as follows.  Bulk gravity
correlators with all points on \im\ are CFT correlators on the
sphere. Inserting additional gravity operators on \ip\ corresponds
to inserting the dual CFT operator at the antipodal point of the
sphere. The reason for this inversion of insertions at \ip\ is
simple. A singularity of a correlator between a point on \im\ and
one on \ip\ can occur only if the two points are connected by a
null geodesic.  A light ray beginning on the sphere at \im\
reaches the antipodal point of the sphere at \ip, as can be easily seen
from figure 2. Therefore 
there is an inversion in the map from the sphere at \ip\ to the
one at \im.\foot{When back reaction is included, the geometry is perturbed, 
and light rays tend to pass the antipodal point \gao. Hence this identification
may be deformed  in perturbation theory. I thank R. Bousso for discussions
of this point.}

This causal connection between points at \ip\ and \im\ has important
consequences for the symmetry group. Naively one might have expected two
copies of the conformal group, one for \ip\ and one for \im, and
correspondingly two 
separate CFTs. However the Green functions know about the 
causal connection between points and therefore transform simply only under
a subgroup of the two conformal groups. The result is a single conformal
group and a single CFT on a
single 
sphere.

We note that only two of the four boundary fields $\phi^\ri_\pm,~~
\phi^\ro_\pm$ are independent, the remaining two being determined by the
equation of motion (at the semiclassical level). These relations are 
the much-studied \desit\ Bogolubov transformations relating \ip\ modes to 
\im\ modes. Therefore there
are at most two independent boundary operators.  

One might alternately have employed one of the other 
Green functions in \rcn\ or \rcnd , such as the Feynman Green function, 
or a Green function associated with one of the other de Sitter invariant
vacua. Such modifications remain to be explored.

\newsec{Quantum States and Virasoro Generators}

A quantum state in the patch ${\cal O}^-$ can be
characterized by its wave function $\Psi$ on the plane \hi. $\Psi$ is a
functional on the space of asymptotically euclidean (on the 
\hi\ plane) two-geometries
$\gamma$. Since the complex diffeomorphisms \dffe\ map this space
to itself, the states $\Psi$ form a representation of the
conformal group. That is, they are states in a conformal field
theory.

The states
$\Psi$ are most naturally described in a radial quantization of the dual 
CFT as
wave functions on the $S^1$ boundary of the \hi. Radial evolution is
generated by $L_0+\bar L_0$. In the bulk description this operator 
generates Killing flow along $z\p_z+\bar z \p_{\bar z}+\p_t$, as depicted
in figure 3.\foot{We note the
absence of a factor of $i$ here. In contrast on the 2D Minkowski cylinder
$L_0+\bar L_0$ generates $i\p_t$. This may be related to the thermal nature
of de Sitter space. } So it
generates 
time evolution
along the planar spacelike slices in \dsmet, accompanied by a dilation. At
large radius the norm of the dilation grows and the Killing vector becomes spacelike.\foot{ An analogy would be the operator $H+J$,
where $H$ is an ordinary Hamiltonian and $J$ a rotation operator.
At large radius the motion generated by this operator is
spacelike, nevertheless it generates evolution along spacelike
slices.} 
The eigenvalue of $L_0+\bar L_0$ is (up to a constant) a conserved
charge known as the AD mass \ad.
\fig{The arrows indicate the direction of the Killing flow generated by 
$L_0+\bar L_0$ within ${\cal O}^-$. Note that it is timelike along the
worldline of an observer at the south pole but becomes spacelike on 
\hi. 
}{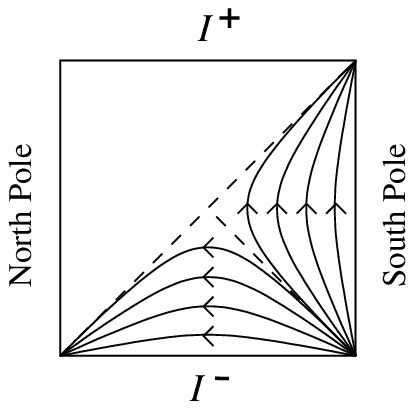}{2in}

As usual, the generators $H(\zeta)$ of any of the diffeomorphisms $\zeta(U)$
can be written as a surface integral at infinity (in this case the
circle $z\bar z \to \infty$) after gauge fixing, imposing  the
constraints and constructing the Dirac brackets. The full expression,
as given in \bh, 
is 
\eqn\rrl{\eqalign{H(\zeta)& ={ 1 \over 16 \pi G} \int dS_\mu \bigl\{  \bigl[\sqrt{\gamma }(
\gamma^{\mu \lambda}\gamma^{\nu \rho}-\gamma^{\mu \nu} \gamma^{\lambda \rho})
(\zeta^t \delta \gamma_{\lambda \rho; \nu}-\zeta^t_{,k} \delta 
\gamma_{\lambda \rho}) \bigr] \cr~~~~~&+2\zeta_\nu \delta \pi^{\mu\nu}
+(2\zeta^\lambda \pi^{\rho \mu}-\zeta^\mu \pi^{\lambda \rho})\delta 
\gamma_{\lambda \rho} \bigr\} .}}
In this expression, $\pi^{\mu\nu}=\sqrt{\gamma }
(K^{\mu\nu}-K\gamma^{\mu\nu})  $ is $16\pi G$ times the momentum conjugate to
$\gamma_{\mu\nu}$, and the prefix $\delta $ denotes the deviation of the 
metric and momentum from their fiducial \ds\ values 
$\gamma_{0 z \bar z}={\ell^2 \over 2} e^{-2t}$  and 
$\pi_0^{z\bar z}=1$. Imposing the boundary conditions \tyip\ and using 
expression \dffe\ for $\zeta$ we find many
terms (including all those in the square parenthesis)
vanish as one approaches  \hi. \rrl\ simplifies to 
\eqn\hhj{H(\zeta)=-{i \over
8\pi G \ell}\int dz \bigl(\zeta^z \pi^{z\bar z}\gamma_{z z}+\zeta^z \gamma_{z
\bar z}\pi^{\bar z\bar z}\bigr).}
In terms of the boundary stress tensor given in \gty, this becomes
\eqn\hxhj{H(\zeta)={1 \over 2 \pi i}\int dz T_{zz}\zeta^z.}
Defining $\zeta_n=\zeta(z^{n+1})$, we have 
\eqn\fcv{L_n\equiv H(\zeta_n)= {1 \over 2 \pi i }\int dz T_{zz}z^{n+1}.}

Expression \fcv\ can also be more directly derived in the 
formalism of \by , where for every boundary symmetry 
$\zeta^\nu$, there is an associated conserved current
$T_{\mu\nu}\zeta^\nu$.
The associated charge is then just the integral of the normal component of
the current around the contour, which is precisely expression \fcv.  

It is tempting to try to compute the de Sitter entropy by 
applying the Cardy formula to these states. This will be explored 
in \rbas.

\newsec{dS$_{D}$/CFT$_{D-1}$ Correspondence}

The \ds /CFT$_2$ correspondence discussed in the preceding sections has an 
obvious generalization to higher dimensions which we briefly mention 
in this section. 
It states that bulk quantum gravity on dS$_D$ is
holographically dual to 
a euclidean conformal field theory on S$^{D-1}$.  
The planar metric for dS$_D$ is 
\eqn\pls{{ds^2 \over \ell^2}=-dt^2+e^{-2t}d \vec x \cdot d \vec x,}
while the spherical metric is 
\eqn\ftvv{{ds^2 \over \ell^2}=-d\tau^2+\cosh^2\tau d\Omega_{D-1}^2,}
with $d\Omega_{D-1}^2$ the unit metric on $S^{D-1}$. 
It can be seen from \ftvv\ that in de Sitter space of any dimension
that a light ray on 
\im\ reaches the antipodal point of the sphere at \ip. Therefore 
the boundary CFT should always involve a single sphere, but the arguments 
of bulk correlators on
\ip\ should map to antipodal points on the sphere, 
relative to those from \im.  One also finds from the 
asymptotic behavior of the wave equation that equation \hmrel\ for the 
conformal weights is generalized to
\eqn\hmreld{h_\pm=\half\bigl( (D-1)\pm\sqrt{(D-1)^2-4m^2\ell^2}\bigr).}
Again we see complex conformal weights for sufficiently massive states. 

 \centerline{\bf Acknowledgements} I am
grateful to J. deBoer, R. Dijkgraaf, S. Giddings, R. Gopakumar, G.
Horowitz, D. Klemm, J. Maldacena, A. Maloney, S. Minwalla, M. Spradlin  
and especially R. Bousso for
useful conversations. This work was supported in part by DOE grant
DE-FG02-91ER40654.

\appendix{A}{\ds\ Geometry}

\ds\ is described by the hyperboloid
\eqn\hpr{X^2+Y^2+Z^2-T^2=\ell^2}
in 3+1 Minkowski space.
The planar coordinates $(z,t)$ are defined by
\eqn\plc{\eqalign{t&=- \ln {Z-T \over \ell},\cr
                  z&={X+iY \over Z-T},\cr
            Z-T&=\ell e^{-{t }},\cr
           Z+T&=\ell e^{{t }}
-\ell {z\bar z } e^{-{t }},\cr X+iY&= \ell z e^{-t } .}}
These lead to the metric \dsmet\ 
\eqn\dsamet{{ds^2 \over \ell^2}=e^{-2t}dzd\bar z -dt^2.}
The spherical coordinates $(\tau,w)$ in \spr\ are defined
by
\eqn\ssrdf{\eqalign{T&=\ell  \sinh{\tau },\cr
           Z&=\ell{1-w\bar w \over 1+ w \bar w} \cosh {\tau },\cr
         X+iY&={2\ell w \cosh {\tau } \over 1+w\bar w}.}}
These give the metric 
\eqn\spr{{ds^2\over\ell^2}
=-d\tau^2+4 \cosh^2{\tau } {dwd\bar w \over (1+w\bar w)^2},}
where $w=\tan {\theta \over 2} e^{i\phi}$ is a complex coordinate
on the round sphere.
The relation between the
spherical coordinates and the planar coordinates  is
\eqn\relt{\eqalign{z&={ w(1+e^{2\tau }) \over 1-w\bar we^{2\tau
}},\cr t&=\tau- \ln \bigl[ {1-w\bar w e^{2\tau }\over1+ w \bar w
}\bigr].}}

The geodesic distance $d(X, X^\prime)$ between two points $X$ and
$X^\prime$ has a simple expression in terms of the Minkowski
coordinates. Define \eqn\sgm{\ell^2
P(X,X^\prime)=XX^\prime+YY^\prime+ZZ^\prime-TT^\prime.} Then
\eqn\dsz{d=\ell \cos^{-1}P.} De Sitter invariance implies that the
Hadamard 
two point function 
\eqn\trf{G(X,X')={\rm const.}<0|\{ \phi(X),\phi(X')\}|0 > }
in a de Sitter invariant vacuum state is a function only of $d$, or
equivalently $P$. Hence away from singularities  $G $ obeys
\eqn\gerq{(P^2-1)\p_P^2G+3P\p_PG+m^2\ell^2 G=0.} 
This equation has two linearly independent solutions 
(related by $P\to -P$), 
corresponding to the existence of a one-parameter family of 
de Sitter invariant vacua.  Among these only one has singularities only
along the light 
cone $P=1$.  This solution is the  hypergeometric function
\eqn\frtk{G(P)={\rm Re}F(h_+,h_-,{3 \over 2};{1+P \over 2}).}

In planar coordinates one finds near \hi :
\eqn\plcf{\lim_{t,t'\to-\infty}P(t,z,\bar z;t', v, \bar v)=-\half
e^{-t-t'}|z-v|^2.}
This diverges, so $G$ can be evaluated with the aid of the 
transformation formula,
\eqn\tsf{F(h_+,h_-,{3 \over 2};z)={\Gamma({3 \over 2}) \Gamma
(h_--h_+) \over  \Gamma(h_-) \Gamma({3 \over 2}-h_+)}(-z)^{-h_+}
F(h_+,h_+-\half,h_++1-h_-;{1 \over z})+(h_+\leftrightarrow h_-)}
and $F(\alpha ,\beta  ,\gamma ;0)=1$. One finds
\eqn\xxvm{\lim_{t,t'\to-\infty}G(t,z,\bar z;t', v, \bar v)={4^{h_+}\Gamma({3 \over 2}) \Gamma
(h_--h_+) \over  \Gamma(h_-) \Gamma({3 \over
2}-h_+)}{e^{h_+(t+t')}\over |z-v|^{2h_+}}
+(h_+\leftrightarrow h_-),}
as given in \geq.

In spherical coordinates one finds near \im\
\eqn\pxcf{\lim_{\tau,\tau'\to-\infty}P(\tau ,w,\bar w;\tau ', v, \bar v)=-
{e^{-\tau-\tau'}|w-v|^2 \over 2(1+w\bar w)(1+ v \bar v)}.}
Using \tsf\ then leads to equation \ftl\ for $G$ on \im. 
We also need $G$ for the case that  $\tau'$ approaches 
\ip\ while $\tau$ approaches \im. This can be deduced from the 
fact that inverting one of the arguments of $P$ (i.e. $X\to -X$) simply 
changes it sign. Hence   
\eqn\zpcf{P(\tau ,w,\bar w;\tau',
v, \bar v)=-P(\tau ,w,\bar w;-\tau', 
-{1 \over \bar v}, -{1 \over v}).}
Hence $P\to +\infty$ for one point on \im\ and one on \ip.
$F$ has a singularity at $P=1$ and a branch cut extending from 
from  $P=1$ to infinity. Using the fact that $G$ is the real part of $F$ 
then yields \rtv . 

\listrefs
\end